\documentclass[aps,prl,reprint,
  superscriptaddress,
  floatfix,
  nofootinbib
]{revtex4-2}

\usepackage{amsmath,amssymb,bm}
\usepackage{mathtools}
\usepackage{graphicx}
\usepackage[hidelinks]{hyperref}
\usepackage{booktabs}
\usepackage{color}

\newcommand{\tr}{\operatorname{tr}}
\newcommand{\dev}{\operatorname{dev}}

\begin{document}

\title{Orientational origin of nonlinear shear rheology in unentangled polymer melts}

\author{Souta Miyamoto}
\email{s.miyamoto@cheme.kyoto-u.ac.jp}
\affiliation{Department of Chemical Science and Engineering, Kyoto University,
Kyoto 615-8510, Japan}

\author{Takeshi Sato}
\affiliation{Advanced Manufacturing Technology Institute, Kanazawa University, Kanazawa 920-1192, Japan}

\author{John J. Molina}
\affiliation{Department of Chemical Science and Engineering, Kyoto University,
Kyoto 615-8510, Japan}

\author{Takashi Taniguchi}
\affiliation{Department of Chemical Science and Engineering, Kyoto University,
Kyoto 615-8510, Japan}

\date{\today}

\begin{abstract}
Universal rheological properties of unentangled polymer melts are investigated through the stretch--orientation decomposition of polymer dynamics. 
By externalizing stretch from the explicit state of each Rouse mode, we construct a constitutive model with orientational kinetics that preserves the Rouse relaxation spectrum.
The resulting model yields the observed steady-shear scalings of the shear viscosity and first normal-stress coefficient, predicts a finite second normal-stress coefficient, and approximately reproduces the Cox--Merz and Gleissle--Osaki rules.
The derived steady--oscillatory shear correspondence further identifies an elastic contribution, associated with the first normal stress difference, that is absent from the conventional Cox--Merz rule. 
\end{abstract}

\maketitle

\emph{Introduction.}--- Long polymer chains exhibit dynamics over broad temporal and spatial scales, and respond to external deformations through a hierarchy of relaxation modes~\cite{Chong2002-lx}.
One of the central questions in polymer physics is whether such coarse-grained modes can provide a unified account of both linear and nonlinear responses~\cite{Wingstrand2015-hm}.
Polymer-melt rheology provides a typical setting for this question.
Mesoscopic models seek to connect chain motion to macroscopic stress responses~\cite{Bird1987-la,Doi1988-cc,Larson1988-nn}.
Specifically, the Rouse model, which represents a polymer chain as Brownian beads connected by Hookean springs, is the canonical example~\cite{Rouse1953-og,Ferry1955-ls,Lodge1971-zc}.

For unentangled polymer melts, the Rouse model captures the characteristic linear viscoelastic response.
It predicts the chain-length ($N$) dependences of the longest mode, namely, the Rouse relaxation time $\lambda_{\mathrm R}\sim N^2$, and the zero shear viscosity $\eta_0\sim N$, together with the Rouse spectrum for the storage and loss moduli: $G'\sim G''\sim\omega^{1/2}$~\cite{Matsumiya2022-df}.
Under steady shear, it predicts shear-rate-independent viscosity $\eta$ and first normal-stress coefficient $\Psi_1$ for shear rate $\dot\gamma$, together with a vanishing second one, $\Psi_2=0$.
By contrast, experiments show shear thinning of $\eta$ and $\Psi_1$ and a finite  $\Psi_2$~\cite{Dalne2025-tx}.
Additionally, steady and oscillatory shear responses, ($\eta$, $\Psi_1$) and ($G'$, $G''$), are empirically related through the Cox--Merz rule~\cite{Cox1958-vy,Venkatraman1990-ys,Winter2009-qt,Snijkers2014-aj} (i.e., $\eta(\dot\gamma)\simeq [G'(\omega)^2+G''(\omega)^2]^{1/2}/\omega$ for $\omega=\dot\gamma$), and the Gleissle--Osaki relation~\cite{Gleissle1980-wb,Osaki1998-bm} (i.e., $\Psi_1(\dot\gamma)\simeq 2G'(\omega)/\omega^2$ for $\omega=\dot\gamma$), which is closely related to Laun-type normal-stress correlations \cite{Laun1986-fo,Das2024-uz}. 
Despite their broad applicability, the common physical origin of these steady--oscillatory correspondences remains unresolved.

Existing explanations introduce additional physical pictures into the Rouse model.
Shear thinning has been explained by finite extensibility~\cite{Peterlin1966-jg}, flow-induced reduction of monomeric friction~\cite{Ianniruberto2012-tf,Yaoita2012-uv,Ianniruberto2020-ap,Miyamoto2023-fy,Psevdos2026-iq}, or shear-dependent length scales, such as tension blobs and shear slits~\cite{Pincus1976-vr,Colby2007-gu,Parisi2021-rs,Dalne2025-tx}.
The Cox--Merz and Gleissle--Osaki relations have been rationalized within integral constitutive frameworks by combining broad relaxation spectra with strain-dependent damping~\cite{Rathinaraj2022-lt}.
Related approaches have also treated bead friction and Brownian force intensity as non-equilibrium parameters~\cite{Watanabe2021-eb,Sato2021-pb}.
These pictures capture different aspects of the observed behavior, but do not identify a common underlying constitutive description.

The standard representation of the stress responses in the Rouse model is a superposition of upper-convected Maxwell (UCM) models.
Each mode is described by an extra-stress tensor or conformation tensor, which combines conformational orientation with stretch.
Some studies have resolved distinct stretch and orientational relaxation processes under deformation~\cite{Marrucci1996-ks,McLeish1998-uk,Marrucci2003-vu,Likhtman2003-ue,O-Connor2018-ee,Lam2018-os,Narimissa2020-ax,Borger2020-ww}.
Following this line, we previously derived an orientation-based constitutive framework from dumbbell dynamics using the deviatoric stress tensor, in which the explicit stretch state is externalized while its contributions remain~\cite{Miyamoto2026implicit}.
This implicit-stretch picture isolates the orientational kinetics of an individual mode.

In this Letter, we propose a constitutive equation derived by applying this stretch--orientation decomposition mode by mode to a Rouse chain.
The stretch contribution is not discarded, but is represented through effective stress scales and relaxation times in the orientational kinetics.
However, even when these effective coefficients are held constant, the orientation-based constitutive equation, combined with the Rouse spectrum, yields the characteristic steady-shear scalings $\eta\sim\dot{\gamma}^{-1/2}$ and
$\Psi_1\sim\dot{\gamma}^{-3/2}$. 
The same orientational description gives the correspondences $\eta(\dot\gamma)\simeq G''(\omega)/\omega$ and $\Psi_1(\dot\gamma)\simeq 2G'(\omega)/\omega^2$ for $\omega\simeq \dot\gamma$, that is, approximately recovers the Cox--Merz and Gleissle--Osaki relations.
Furthermore, it identifies an elastic contribution to the steady--oscillatory correspondence, carried by the first normal stress difference and absent from the conventional Cox--Merz rule.

\begin{figure}[htbp]
    \centering
    \includegraphics[width=0.9\linewidth]{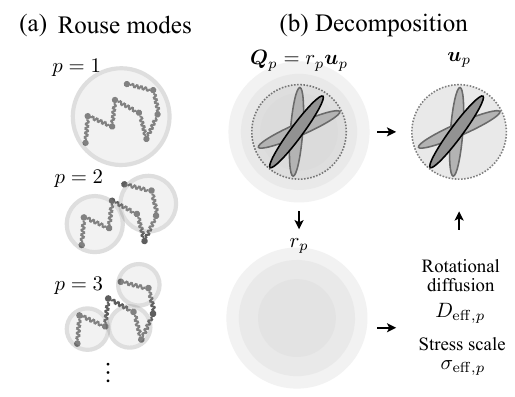}
    \caption{Schematic illustration of (a) Rouse modes and (b) the stretch--orientation decomposition of each mode.}
    \label{fig:mode_decomposition}
\end{figure}

\emph{Rouse model.}--- The Rouse model describes a linear chain as $N$ segments, with large $N$ considered here.
The dynamics of a Rouse chain can be decomposed into independent modes on the normal coordinates (Fig.~\ref{fig:mode_decomposition}a), where each mode follows dumbbell dynamics, represented by two beads connected by a Hookean spring.
Correspondingly, the extra stress tensor ($\bm\tau$) is expressed as the sum of the stress contributions from each mode ($\bm\tau_p$), each following the UCM model:
\begin{align}
    &\bm\tau = \sum_{p=1}^N \bm\tau_p;\quad \overset{\triangledown}{\bm\tau}_p + \frac{1}{\lambda_p}\bm\tau_p = 2G_p\bm D,
    \label{eq:UCM-Rouse}
\end{align}
where $\lambda_p$ and $G_p$ are the $p$-th mode relaxation time and modulus, respectively, and $\bm D$ is the deformation rate tensor.
$\triangledown$ placed over $\bm\tau_p$ denotes the upper-convected time derivative of $\bm\tau_p$.
From the standard analysis~\cite{Bird1987-la,Doi1988-cc}, we obtain the characteristic quantities:
\begin{align}
    \lambda_p = \frac{\lambda_\mathrm{R}}{p^2},\quad \lambda_\mathrm{R}=\frac{N^2\zeta}{2\pi^2h},\quad G_p=G=\nu k_\mathrm{B}T,
\end{align}
where $\zeta$ is the friction coefficient,  $h$ the spring constant,  $\nu$ the chain number density,  $k_\mathrm{B}$ the Boltzmann constant, and $T$ the temperature.

In the Rouse regime between the longest and shortest relaxation-time scales, the superposed relaxation modulus, $G(t) = \sum_{p=1}^N G_p\exp(-t/\lambda_p)$, follows the scaling $G(t)\sim t^{-1/2}$. 
The corresponding frequency-domain scalings of the linear storage and loss moduli are $G'(\omega)\sim \omega^{1/2}$ and $G''(\omega)\sim \omega^{1/2}$.
At a small strain amplitude, $G'(\omega)$ and $G''(\omega)$ are
\begin{align}
    G'(\omega)  = \sum_{p=1}^N \frac{G_p(\lambda_p\omega)^2}{1+(\lambda_p\omega)^2},\ 
    G''(\omega) = \sum_{p=1}^N \frac{G_p\lambda_p\omega}{1+(\lambda_p\omega)^2}.
    \label{eq:oscillatory}
\end{align}
Based on Eq.~\eqref{eq:UCM-Rouse}, we obtain the steady shear responses:
$\sigma_\mathrm{s}(\dot\gamma)=\sum_{p=1}^N G_p\lambda_p\dot\gamma$,
$N_1(\dot\gamma)=\sum_{p=1}^N 2G_p(\lambda_p\dot\gamma)^2$,
$N_2(\dot\gamma)=0$,
where $\sigma_\mathrm{s}$ is the shear stress, and $N_1$ and $N_2$ are the first and second normal stress differences, respectively.
Thus, the constant shear viscosity and normal stress coefficients are 
\begin{equation}
    \eta(\dot\gamma)=\sum_{p=1}^N G_p\lambda_p,\ \Psi_1(\dot\gamma) = \sum_{p=1}^N 2G_p\lambda_p^2,\ \Psi_2(\dot\gamma)=0.
\end{equation}
However, experiments show deviations at higher shear rates $\dot\gamma > 1/\lambda_\mathrm{R}$, in the form of shear thinning.
The Rouse model provides the equivalence between steady and oscillatory shear responses in the linear regime, whereas the empirical Cox--Merz and Gleissle--Osaki relations remain approximately applicable well into the nonlinear regime. 
The origin of both the nonlinear steady-shear response and its correspondence with the oscillatory response remains unresolved.

\begin{figure*}[htbp]
    \centering
    \includegraphics[width=1.0\linewidth]{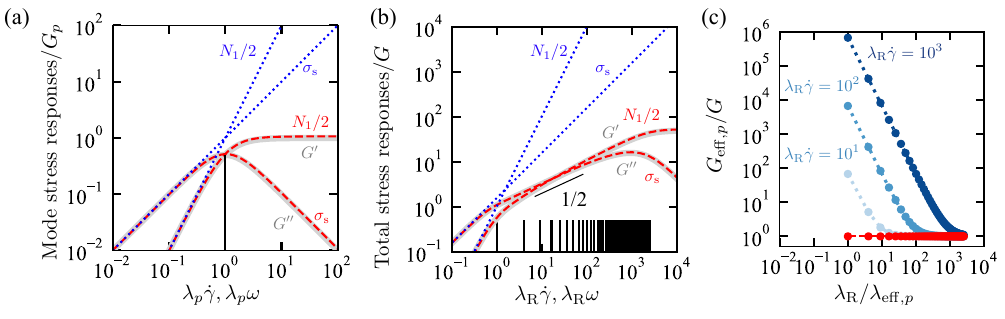}
    \caption{Stress responses ($N=50$) for (a) a single mode $p$ and (b) the total response, and (c) the corresponding effective coefficients. Thick gray solid lines show the responses evaluated from small-amplitude oscillatory shear data. Blue dotted lines show the steady shear responses from the conventional UCM model and their superposition. 
    Red dashed lines are those from the OIS model. Vertical lines correspond to the relaxation modes. In (c), the curved lines are obtained from the UCM superposition, whereas the horizontal line corresponds to the OIS model with constant effective coefficients. 
    }
    \label{fig:rouse_scalings}
\end{figure*}

\emph{Orientational constitutive equation.}--- Based on the standard Rouse description, additional mechanisms for mobility reduction under deformation have been proposed to predict the shear-thinning exponents.
In these mechanisms, the orientational states of the chains play important roles, and are correlated with the mobility.
However, the stress state $\bm\tau$ contains mixed contributions from both stretch and orientation, and $\tr\bm\tau$ particularly reflects the stretch state of the chain conformation. 
Thus, we previously proposed a constitutive description based on orientational kinetics in the dumbbell model, using the deviatoric stress tensor $\bm s$, such that $\tr\bm s=0$~\cite{Miyamoto2026implicit}.
The stretch--orientation decomposition for the dumbbell model focuses on the orientational contributions without explicit stretch states. 

Applying this decomposition mode by mode gives an orientational constitutive equation for the Rouse modes (Fig.~\ref{fig:mode_decomposition}b).
Integrating out the radial coordinate externalizes the explicit stretch degree of freedom while retaining its effects through two effective coefficients for the orientational kinetics: a stress scale $\sigma_{\mathrm{eff},p}$ and a rotational diffusivity $D_{\mathrm{eff},p}$.
These coefficients enter the constitutive equation through the effective relaxation modulus $G_{\mathrm{eff},p}=\sigma_{\mathrm{eff},p}/5$ and relaxation time $\lambda_{\mathrm{eff},p} = 1/\left(6D_{\mathrm{eff},p}- {d\ln\sigma_{\mathrm{eff},p}}/{dt}\right)$ (see Ref.~\cite{Miyamoto2026implicit}).
The resultant equation for the mode-wise deviatoric stress tensor is
\begin{align}
    &\overset{\triangledown_0}{\bm s_p}
    +\frac{1}{\lambda_{\mathrm{eff},p}}\bm s_p    
    +\xi\dev(\bm s_p\cdot\bm D+\bm D\cdot\bm s_p)
    = 2G_{\mathrm{eff},p}\bm D.
    \label{eq:OIS-model}
\end{align}
Here, $\triangledown_0$ placed over $\bm s_p$ means the traceless upper convected derivative of $\bm s_p$, by adding $2(\bm s_p:D)\mathbf{I}/3$.
The term $\dev(\bm s_p\cdot\bm D+\bm D\cdot\bm s_p)$ originates from affine deformation. 
At first order in the expansion about the isotropic orientational state, $\xi=4/7$~\cite{Miyamoto2026implicit}.
We hereafter refer to Eq.~\eqref{eq:OIS-model} as the orientation-with-implicit-stretch (OIS) model.

The stress responses of the OIS model under steady shear are given as
\begin{align}
    \sigma_\mathrm{s}(\dot\gamma) &= \sum_{p=1}^N\frac{G_{\mathrm{eff},p}\lambda_{\mathrm{eff},p}\dot\gamma}{1+K_\xi(\lambda_{\mathrm{eff},p}\dot\gamma)^2},
    \label{eq:eff-steady-shear-1}\\
    N_1(\dot\gamma)    &= \sum_{p=1}^N\frac{2G_{\mathrm{eff},p}(\lambda_{\mathrm{eff},p}\dot\gamma)^2}{1+K_\xi(\lambda_{\mathrm{eff},p}\dot\gamma)^2},\ N_2(\dot\gamma)    = -\frac{\xi}{2}N_1,   
    \label{eq:eff-steady-shear-2}
\end{align}
where $K_\xi = 1-(1-\xi)^2/3$.
$K_\xi$ remains finite even at $\xi=0$, and thus the denominator is proportional to $(\lambda_\mathrm{eff}\dot\gamma)^2$ at high shear rates. 
In general, these effective coefficients may depend on the steady-shear state. 
Here, we instead hold them constant.
When held constant, consistency with the linear Rouse response requires $\lambda_{\mathrm{eff},p}=\lambda_p$ and $G_{\mathrm{eff},p}=G_p$.
This choice leaves a bounded orientational anisotropy whose progressive alignment with the shear flow direction produces shear-stress turnover and normal-stress saturation for each mode.
The small-amplitude oscillatory response is identical to that of the original Rouse model, because the nonlinear terms are of second order in the deformation amplitude.

\emph{Exponents for steady shear.}--- Figure~\ref{fig:rouse_scalings} shows how mode-wise orientational saturation combined with the Rouse spectrum yields the steady-shear scaling behavior.
In Fig.~\ref{fig:rouse_scalings}(a), the $p$-th mode gives a turnover followed by a decrease in $\sigma_{\mathrm{s},p}(\dot\gamma)$ and saturation of $N_{1,p}(\dot\gamma)/2$, corresponding to $G_p''(\omega)$ and $G_p'(\omega)$, respectively.
In Fig.~\ref{fig:rouse_scalings}(b), the stress response of a Rouse chain for $\sigma_\mathrm{s}(\dot\gamma)$ and $N_1(\dot\gamma)$ exhibits the scaling $\dot\gamma^{1/2}$, corresponding to the Rouse spectrum $G'(\omega),G''(\omega)\sim\omega^{1/2}$ for the Rouse regime.
This correspondence between steady and oscillatory shear flows follows directly from the analogous expressions in Eqs.~\eqref{eq:oscillatory}, \eqref{eq:eff-steady-shear-1}, and \eqref{eq:eff-steady-shear-2}.
In Fig.~\ref{fig:rouse_scalings}(c), we show that the conventional UCM model exhibits an increasing effective relaxation modulus, which maintains the monotonic increases $\sigma_\mathrm{s}\sim\dot\gamma$ and $N_1\sim\dot\gamma^2$. 
Holding the effective coefficients constant removes this stretch-related contribution.
Thus, although affine orientation produces strong thinning at the single mode, superposition over the Rouse spectrum converts the mode-wise saturation into the observed power-law response.

\begin{figure*}[htbp]
    \centering
    \includegraphics[width=1.0\linewidth]{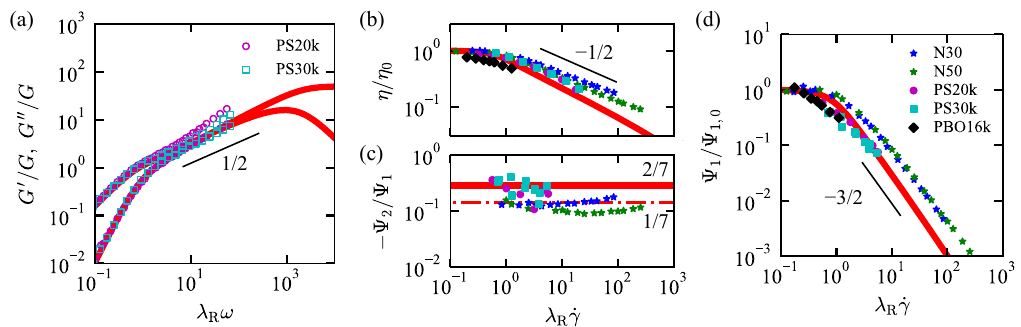}
    \caption{Comparison with experimental results for PS20k and PS30k from Dalne \textit{et al.}~\cite{Dalne2025-tx}, PBO16k from Matsumiya \textit{et al.}~\cite{Matsumiya2022-df}, and Kremer--Grest model results for N30 and N50 from Xu \textit{et al.}~\cite{Xu2014-wh}.
    Open and filled symbols denote oscillatory and steady-shear data, respectively.
    Red lines show predictions by the OIS model with $N=50$.
    Panels show (a) the storage and loss moduli, (b) the steady shear viscosity, (c) the ratio of the second to first normal stress difference coefficients, and (d) the first normal stress difference coefficient.
    In (c), the solid and dash-dotted lines correspond to $-\Psi_2/\Psi_1=2/7$ and $1/7$, respectively, associated with $\xi=4/7$ and $2/7$.
    }
    \label{fig:comparison}
\end{figure*}

\emph{Material functions under shear.}--- We can obtain the following scaling laws for material functions in the Rouse regime: $\eta \sim \dot\gamma^{-1/2}$, $\Psi_1 \sim \dot\gamma^{-3/2}$, $\Psi_2/\Psi_1=-2/7$.
The OIS model accounts for these shear material functions within a single framework, without introducing an explicit flow-dependent length scale.
Figure~\ref{fig:comparison} assesses the theoretical treatment against unentangled polystyrene (PS) melts~\cite{Dalne2025-tx}, poly(butylene oxide) (PBO) melts~\cite{Matsumiya2022-df}, and Kremer--Grest (KG) simulations~\cite{Kremer1990-xa,Xu2014-wh} with unentangled chains.
In Fig.~\ref{fig:comparison}(a), the Rouse spectrum is captured by the relaxation-time hierarchy, except for the discrepancy in the glassy high-frequency regime.
In Fig.~\ref{fig:comparison}(b)--(d), we confirm the scaling behavior of the material functions.
The known scalings, $\eta\sim\dot\gamma^{-1/2}$ and $\Psi_1\sim\dot\gamma^{-3/2}$, are reproduced, and $-\Psi_2/\Psi_1=2/7$ is consistent with the experimental PS results, but not with the KG simulations.
The agreement of $\eta$ and $\Psi_1$ supports orientational mode saturation as a common mechanism, while the residual variation in $\Psi_2$ may reflect sensitivity to microscopic many-chain or intersegment correlations.

\begin{figure}
    \centering
    \includegraphics[width=0.95\linewidth]{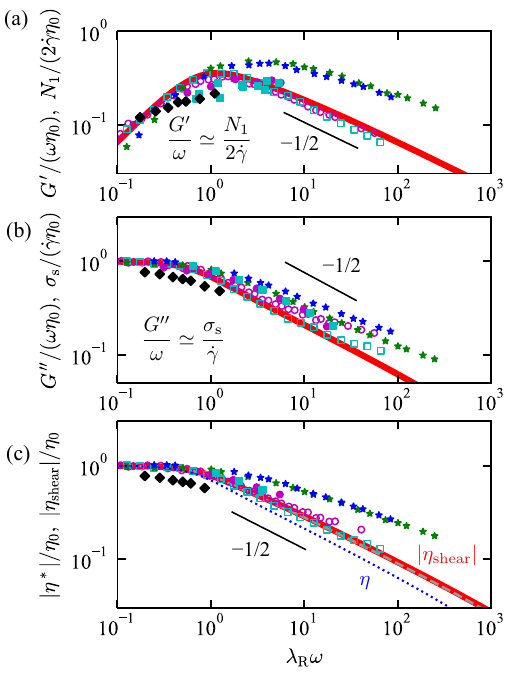}
    \caption{Steady--oscillatory shear correspondences for (a) $G'/\omega$ and $N_1/(2\dot\gamma)$, (b) $G''/\omega$ and $\sigma_\mathrm{s}/\dot\gamma$, and (c) the norm of the steady-shear viscosity pair. In (c), the blue dotted line represents the shear viscosity $\eta$. Symbols are the same as in Fig.~\ref{fig:comparison}.
    }
    \label{fig:steady_oscillatory}
\end{figure}
\emph{The generalized Cox--Merz rule.}--- For the present constant-coefficient orientational model, Eqs.~\eqref{eq:oscillatory}, \eqref{eq:eff-steady-shear-1}, and \eqref{eq:eff-steady-shear-2} give
\begin{align}
    \frac{\sigma_\mathrm{s}}{\dot\gamma} = \frac{G''}{\omega}\Bigr|_{\omega=\sqrt{K_\xi}\dot\gamma},\quad 
    \sqrt{K_\xi}\frac{N_1}{2\dot\gamma} = \frac{G'}{\omega}\Bigr|_{\omega=\sqrt{K_\xi}\dot\gamma}.
\end{align}
Thus, the generalized Cox--Merz rule is also derived:
\begin{equation}
    \left\{[\eta(\dot\gamma)]^2
    +K_\xi\left[\frac{N_1(\dot\gamma)}{2\dot\gamma}\right]^2\right\}^{1/2} 
    =
    |\eta^*(\omega)|\Bigr|_{\omega=\sqrt{K_\xi}\dot\gamma}.
    \label{eq:general-Cox-Merz}
\end{equation}
Neglecting the elastic contribution $N_1/(2\dot\gamma)$ on the left-hand side reduces Eq.~\eqref{eq:general-Cox-Merz} to the conventional Cox--Merz relation, apart from the factor $K_\xi$.
For the orientational value $\xi = {4}/{7}$, $K_\xi = {46}/{49} = 0.939...$, and $\sqrt{K_\xi}=0.968...$, so the frequency rescaling is small. 
Thus, we use $K_\xi\simeq1$ in the following comparison.
The relation between $N_1$ and $G'$ corresponds to the Gleissle--Osaki relation, and is closely related to Laun-type normal-stress correlations.
These formulas identify the elastic contribution missing from the Cox--Merz rule, namely, the steady-shear counterpart of the oscillatory quantity $G'/\omega$.
In summary, we can define the following steady-shear viscosity pair:
\begin{equation}
    \eta_\mathrm{shear}=\left(\frac{N_1}{2\dot\gamma},\frac{\sigma_\mathrm{s}}{\dot\gamma}\right)
    \simeq
    \left(\frac{G'}{\omega},\frac{G''}{\omega}\right)\Bigr|_{\omega=\dot\gamma}.
\end{equation}
The orientational kinetics under high-rate shear yield the correspondences between steady and oscillatory shear responses.

Figure~\ref{fig:steady_oscillatory} shows the correspondences between $G'$ and $N_1$ and between $G''$ and $\sigma_\mathrm{s}$. 
Within the Rouse regime, the corresponding quantities exhibit the same $-1/2$ exponent, while neglecting $N_1/(2\dot\gamma)$ produces the difference between $|\eta_\mathrm{shear}|$ and $\eta$ shown in Fig.~\ref{fig:steady_oscillatory}(c).
This helps explain the success of the empirical Cox--Merz rule when $N_1$ is omitted, and this simplification is also experimentally convenient given the difficulty of measuring $N_1$.
Nevertheless, for general materials, the validity of the Cox--Merz rule should be reconsidered from the viewpoint of this elastic contribution.
Note that the slight underestimation of the KG results may indicate a larger stretch contribution in the simulations than in the experimental systems.

\emph{Conclusion.---} We have derived the nonlinear shear rheology of unentangled polymer melts using an orientational constitutive equation with the Rouse modes. 
The resultant equation preserves the linear response, predicts the steady-shear scalings $\eta \sim \dot{\gamma}^{-1/2}$ and $\Psi_1 \sim \dot{\gamma}^{-3/2}$, and provides a unified description of the Cox--Merz and Gleissle--Osaki rules for steady and oscillatory shear. 
It also identifies the first-normal-stress contribution absent from the conventional Cox--Merz rule.
The agreement of the predicted $\eta$ and $\Psi_1$ with experiments and many-chain simulations supports orientational kinetics as an origin of these characteristic nonlinear scalings.
The proposed orientational framework can be extended to other viscoelastic materials.

\emph{Acknowledgments.---}
This work was supported by the Japan Society for the Promotion of Science (JSPS) KAKENHI under Grant No.~25K23525 awarded to S.M. and 26K01281 to T.S. OpenAI ChatGPT was used to assist with parts of the mathematical derivations and to polish the English text. 
The authors provided the physical assumptions and mathematical formulation and independently verified all AI-assisted derivations and reviewed the resulting text.

\bibliographystyle{apsrev4-2}
\bibliography{main}

@misc{Miyamoto2026implicit,
  title         = {Polymer viscoelasticity from an orientational closure with implicit stretch},
  author        = {Miyamoto, Souta and Sato, Takeshi and Jin, Wanxiang
                   and Tanabe, Katsuaki and Molina, John J. and Taniguchi, Takashi},
  year          = {2026},
  eprint        = {2610.00892},
  archivePrefix = {arXiv},
  primaryClass  = {cond-mat.soft},
  doi           = {10.48550/arXiv.2610.00892},
}

@ARTICLE{Chong2002-lx,
  title     = "Mode-coupling theory for structural and conformational dynamics
               of polymer melts",
  author    = "Chong, S-H and Fuchs, M",
  journal   = "Phys. Rev. Lett.",
  publisher = "American Physical Society (APS)",
  volume    =  88,
  number    =  18,
  pages     =  185702,
  year      =  2002,
  doi       = "10.1103/PhysRevLett.88.185702",
}

@ARTICLE{Borger2020-ww,
  title     = "Stretch and orientational mode decoupling in relaxation of highly
               stretched polymer melts",
  author    = "Borger, Anine L and Huang, Qian and Hassager, Ole and
               Kirkensgaard, Jacob J K and Almdal, Kristoffer and Mortensen,
               Kell",
  journal   = "Phys. Rev. Res.",
  publisher = "American Physical Society (APS)",
  volume    =  2,
  number    =  4,
  pages     =  043119,
  year      =  2020,
  doi       = "10.1103/physrevresearch.2.043119",
}

@ARTICLE{Lam2018-os,
  title    = "Scaling behavior of anisotropy relaxation in deformed polymers",
  author   = "Lam, Christopher N and Xu, Wen-Sheng and Chen, Wei-Ren and Wang,
              Zhe and Stanley, Christopher B and Carrillo, Jan-Michael Y and
              Uhrig, David and Wang, Weiyu and Hong, Kunlun and Liu, Yun and
              Porcar, Lionel and Do, Changwoo and Smith, Gregory S and Sumpter,
              Bobby G and Wang, Yangyang",
  journal  = "Phys. Rev. Lett.",
  volume   =  121,
  number   =  11,
  pages    =  117801,
  year     =  2018,
  doi      = "10.1103/PhysRevLett.121.117801",
}

@ARTICLE{Wingstrand2015-hm,
  title     = "Linear and nonlinear universality in the rheology of polymer
               melts and solutions",
  author    = "Wingstrand, Sara L and Alvarez, Nicolas J and Huang, Qian and
               Hassager, Ole",
  journal   = "Phys. Rev. Lett.",
  publisher = "American Physical Society (APS)",
  volume    =  115,
  number    =  7,
  pages     =  078302,
  year      =  2015,
  doi       = "10.1103/PhysRevLett.115.078302",
}

@ARTICLE{O-Connor2018-ee,
  title     = "Relating chain conformations to extensional stress in entangled
               polymer melts",
  author    = "O'Connor, Thomas C and Alvarez, Nicolas J and Robbins, Mark O",
  journal   = "Phys. Rev. Lett.",
  publisher = "American Physical Society (APS)",
  volume    =  121,
  number    =  4,
  pages     =  047801,
  year      =  2018,
  doi       = "10.1103/PhysRevLett.121.047801",
}

@ARTICLE{Xu2014-wh,
  title     = "Shear thinning behavior of linear polymer melts under shear flow
               via nonequilibrium molecular dynamics",
  author    = "Xu, Xiaolei and Chen, Jizhong and An, Lijia",
  journal   = "J. Chem. Phys.",
  publisher = "AIP Publishing",
  volume    =  140,
  number    =  17,
  pages     =  174902,
  year      =  2014,
  doi       = "10.1063/1.4873709",
}

@ARTICLE{Kremer1990-xa,
  title     = "Dynamics of entangled linear polymer melts: A molecular-dynamics
               simulation",
  author    = "Kremer, Kurt and Grest, Gary S",
  journal   = "J. Chem. Phys.",
  publisher = "AIP Publishing",
  volume    =  92,
  number    =  8,
  pages     = "5057--5086",
  year      =  1990,
  doi       = "10.1063/1.458541",
}

@ARTICLE{Narimissa2020-ax,
  title     = "Modeling nonlinear rheology of unentangled polymer melts based on
               a single integral constitutive equation",
  author    = "Narimissa, Esmaeil and Wagner, Manfred H",
  journal   = "J. Rheol.",
  publisher = "Society of Rheology",
  volume    =  64,
  number    =  1,
  pages     = "129--140",
  year      =  2020,
  doi       = "10.1122/1.5128295",
}

@ARTICLE{Likhtman2003-ue,
  title     = "Simple constitutive equation for linear polymer melts derived
               from molecular theory: Rolie--Poly equation",
  author    = "Likhtman, Alexei E and Graham, Richard S",
  journal   = "J. Nonnewton. Fluid Mech.",
  publisher = "Elsevier BV",
  volume    =  114,
  number    =  1,
  pages     = "1--12",
  year      =  2003,
  doi       = "10.1016/s0377-0257(03)00114-9",
}

@ARTICLE{Marrucci2003-vu,
  title     = "Flow-induced orientation and stretching of entangled polymers",
  author    = "Marrucci, Giuseppe and Ianniruberto, Giovanni",
  journal   = "Philos. Trans. A Math. Phys. Eng. Sci.",
  publisher = "The Royal Society",
  volume    =  361,
  number    =  1805,
  pages     = "677--87; discussion 687--8",
  year      =  2003,
  doi       = "10.1098/rsta.2002.1156",
}

@ARTICLE{McLeish1998-uk,
  title     = "Molecular constitutive equations for a class of branched
               polymers: The pom-pom polymer",
  author    = "McLeish, T C B and Larson, R G",
  journal   = "J. Rheol.",
  publisher = "Society of Rheology",
  volume    =  42,
  number    =  1,
  pages     = "81--110",
  year      =  1998,
  doi       = "10.1122/1.550933"
}

@ARTICLE{Lodge1971-zc,
  title     = "Constitutive equations for polymer solutions derived from the
               bead/spring model of Rouse and Zimm",
  author    = "Lodge, A S and Wu, Yeen-Jing",
  journal   = "Rheol. Acta",
  publisher = "Springer Nature",
  volume    =  10,
  number    =  4,
  pages     = "539--553",
  year      =  1971,
  doi       = "10.1007/bf03396404"
}

@ARTICLE{Marrucci1996-ks,
  title     = "Dynamics of entanglements: A nonlinear model consistent with the
               Cox-Merz rule",
  author    = "Marrucci, G",
  journal   = "J. Nonnewton. Fluid Mech.",
  publisher = "Elsevier BV",
  volume    =  62,
  number    = "2-3",
  pages     = "279--289",
  year      =  1996,
  doi       = "10.1016/0377-0257(95)01407-1",
}

@ARTICLE{Sato2021-pb,
  title     = "A constitutive equation for Rouse model modified for variations
               of spring stiffness, bead friction, and Brownian force intensity
               under flow",
  author    = "Sato, Takeshi and Kwon, Youngdon and Matsumiya, Yumi and
               Watanabe, Hiroshi",
  journal   = "Phys. Fluids",
  publisher = "AIP Publishing",
  volume    =  33,
  number    =  6,
  pages     =  063106,
  year      =  2021,
  doi       = "10.1063/5.0055559",
}

@ARTICLE{Peterlin1966-jg,
  title     = "Hydrodynamics of macromolecules in a velocity field with
               longitudinal gradient",
  author    = "Peterlin, A",
  journal   = "J. Polym. Sci. B",
  publisher = "Wiley",
  volume    =  4,
  number    =  4,
  pages     = "287--291",
  year      =  1966,
  doi       = "10.1002/pol.1966.110040411",
}

@INPROCEEDINGS{Gleissle1980-wb,
  author    = {Gleissle, Wolfgang},
  title     = {Two simple time--shear rate relations combining viscosity and
               first normal stress coefficient in the linear and non-linear
               flow range},
  booktitle = {Rheology: Volume 2: Fluids},
  editor    = {Astarita, Giovanni and Marrucci, Giuseppe and Nicolais, Luigi},
  pages     = {457--462},
  publisher = {Springer},
  address   = {Boston, MA},
  year      = {1980},
  doi       = {10.1007/978-1-4684-3743-0_85},
  isbn      = {978-1-4684-3743-0},
}

@ARTICLE{Snijkers2014-aj,
  title     = "Appraisal of the Cox-Merz rule for well-characterized entangled
               linear and branched polymers",
  author    = "Snijkers, Frank and Vlassopoulos, Dimitris",
  journal   = "Rheol. Acta",
  publisher = "Springer Science and Business Media LLC",
  volume    =  53,
  number    =  12,
  pages     = "935--946",
  year      =  2014,
  doi       = "10.1007/s00397-014-0799-6",
}

@ARTICLE{Das2024-uz,
  title     = "Laun's rule for predicting the first normal stress coefficient in
               complex fluids: A comprehensive investigation using fractional
               calculus",
  author    = "Das, Mohua and Rathinaraj, Joshua David John and Palade, Liviu
               Iulian and McKinley, Gareth H",
  journal   = "Phys. Fluids",
  publisher = "AIP Publishing",
  volume    =  36,
  number    =  1,
  pages     =  013111,
  year      =  2024,
  doi       = "10.1063/5.0179709",
}

@ARTICLE{Watanabe2021-eb,
  title     = "Revisiting nonlinear flow behavior of rouse chain: Roles of
               {FENE}, friction-reduction, and Brownian force intensity
               variation",
  author    = "Watanabe, Hiroshi and Matsumiya, Yumi and Sato, Takeshi",
  journal   = "Macromolecules",
  publisher = "American Chemical Society (ACS)",
  volume    =  54,
  number    =  8,
  pages     = "3700--3715",
  year      =  2021,
  doi       = "10.1021/acs.macromol.1c00013",
}

@ARTICLE{Matsumiya2022-df,
  title     = "Rheo-dielectric behavior of unentangled poly(butylene oxide)
               under steady shear: Preliminary evaluation of non-equilibrium
               parameters at the onset of nonlinearity",
  author    = "Matsumiya, Yumi and Sato, Takeshi and Chen, Quan and Watanabe,
               Hiroshi",
  journal   = "Nihon Reoroji Gakkaishi",
  publisher = "Society of Rheology Japan",
  volume    =  50,
  number    =  5,
  pages     = "371--385",
  year      =  2022,
  doi       = "10.1678/rheology.50.371",
}

@ARTICLE{Ferry1955-ls,
  title     = "Extensions of the Rouse theory of viscoelastic properties to
               undiluted linear polymers",
  author    = "Ferry, John D and Landel, Robert F and Williams, Malcolm L",
  journal   = "J. Appl. Phys.",
  publisher = "AIP Publishing",
  volume    =  26,
  number    =  4,
  pages     = "359--362",
  year      =  1955,
  doi       = "10.1063/1.1721997",
}

@ARTICLE{Osaki1998-bm,
  title     = "On the estimation of stresses in steady shear flow from the
               dynamic viscoelasticity for polymeric liquids",
  author    = "Osaki, K and Watanabe, H and Inoue, T",
  journal   = "Nihon Reoroji Gakkaishi",
  publisher = "Society of Rheology Japan",
  volume    =  26,
  number    =  1,
  pages     = "49--52",
  year      =  1998,
  doi       = "10.1678/rheology1973.26.1\_49"
}

@ARTICLE{Laun1986-fo,
  title     = "Prediction of elastic strains of polymer melts in shear and
               elongation",
  author    = "Laun, H M",
  journal   = "J. Rheol.",
  publisher = "Society of Rheology",
  volume    =  30,
  number    =  3,
  pages     = "459--501",
  year      =  1986,
  doi       = "10.1122/1.549855",
}

@ARTICLE{Winter2009-qt,
  title     = "Three views of viscoelasticity for Cox--Merz materials",
  author    = "Winter, H Henning",
  journal   = "Rheol. Acta",
  publisher = "Springer Science and Business Media LLC",
  volume    =  48,
  number    =  3,
  pages     = "241--243",
  year      =  2009,
  doi       = "10.1007/s00397-008-0329-5",
}

@ARTICLE{Venkatraman1990-ys,
  title     = "A comparison of torsional and capillary rheometry for polymer
               melts: The Cox-Merz rule revisited",
  author    = "Venkatraman, S and Okano, M",
  journal   = "Polym. Eng. Sci.",
  publisher = "Wiley",
  volume    =  30,
  number    =  5,
  pages     = "308--313",
  year      =  1990,
  doi       = "10.1002/pen.760300508",
}

@ARTICLE{Rathinaraj2022-lt,
  title     = "Why the Cox--Merz rule and Gleissle mirror relation work: A
               quantitative analysis using the Wagner integral framework with a
               fractional Maxwell kernel",
  author    = "Rathinaraj, Joshua David John and Keshavarz, Bavand and McKinley,
               Gareth H",
  journal   = "Phys. Fluids",
  publisher = "AIP Publishing",
  volume    =  34,
  number    =  3,
  pages     =  033106,
  year      =  2022,
  doi       = "10.1063/5.0084478",
}

@ARTICLE{Pincus1976-vr,
  title     = "Excluded volume effects and stretched polymer chains",
  author    = "Pincus, P",
  journal   = "Macromolecules",
  publisher = "American Chemical Society (ACS)",
  volume    =  9,
  number    =  3,
  pages     = "386--388",
  year      =  1976,
  doi       = "10.1021/ma60051a002",
}

@ARTICLE{Parisi2021-rs,
  title     = "Nonlinear shear rheology of entangled polymer rings",
  author    = "Parisi, Daniele and Costanzo, Salvatore and Jeong, Youncheol and
               Ahn, Junyoung and Chang, Taihyun and Vlassopoulos, Dimitris and
               Halverson, Jonathan D and Kremer, Kurt and Ge, Ting and
               Rubinstein, Michael and Grest, Gary S and Srinin, Watee and
               Grosberg, Alexander Y",
  journal   = "Macromolecules",
  publisher = "American Chemical Society (ACS)",
  volume    =  54,
  number    =  6,
  pages     = "2811--2827",
  year      =  2021,
  doi       = "10.1021/acs.macromol.0c02839",
}

@ARTICLE{Colby2007-gu,
  title     = "Shear thinning of unentangled flexible polymer liquids",
  author    = "Colby, R H and Boris, D C and Krause, W E and Dou, S",
  journal   = "Rheol. Acta",
  publisher = "Springer Science and Business Media LLC",
  volume    =  46,
  number    =  5,
  pages     = "569--575",
  year      =  2007,
  doi       = "10.1007/s00397-006-0142-y",
}

@ARTICLE{Dalne2025-tx,
  title     = "Nonlinear shear rheology of unentangled polymer melts",
  author    = "Dalne, Maxime and Costanzo, Salvatore and Peponaki, Katerina and
               Jiang, Nuofei and Alexandris, Stelios and Grizzuti, Nino and
               Vlassopoulos, Dimitris and Rubinstein, Michael and van Ruymbeke,
               Evelyne",
  journal   = "Macromolecules",
  publisher = "American Chemical Society (ACS)",
  volume    =  58,
  number    =  14,
  pages     = "7062--7083",
  year      =  2025,
  doi       = "10.1021/acs.macromol.5c00553",
}

@ARTICLE{Miyamoto2023-fy,
  title     = "Stretch-orientation-induced reduction of friction in
               well-entangled bidisperse blends: a dual slip-link simulation
               study",
  author    = "Miyamoto, Souta and Sato, Takeshi and Taniguchi, Takashi",
  journal   = "Rheol. Acta",
  publisher = "Springer Science and Business Media LLC",
  volume    =  62,
  number    =  1,
  pages     = "57--70",
  year      =  2023,
  doi       = "10.1007/s00397-022-01378-5",
}

@ARTICLE{Ianniruberto2012-tf,
  title     = "Simulations of fast shear flows of {PS} oligomers confirm
               monomeric friction reduction in fast elongational flows of
               monodisperse {PS} melts as indicated by rheooptical data",
  author    = "Ianniruberto, G and Brasiello, A and Marrucci, G",
  journal   = "Macromolecules",
  publisher = "American Chemical Society (ACS)",
  volume    =  45,
  number    =  19,
  pages     = "8058--8066",
  year      =  2012,
  doi       = "10.1021/ma301368d",
}

@ARTICLE{Yaoita2012-uv,
  title     = "Primitive chain network simulation of elongational flows of
               entangled linear chains: Stretch/orientation-induced reduction of
               monomeric friction",
  author    = "Yaoita, Takatoshi and Isaki, Takeharu and Masubuchi, Yuichi and
               Watanabe, Hiroshi and Ianniruberto, Giovanni and Marrucci,
               Giuseppe",
  journal   = "Macromolecules",
  publisher = "American Chemical Society (ACS)",
  volume    =  45,
  number    =  6,
  pages     = "2773--2782",
  year      =  2012,
  doi       = "10.1021/ma202525v",
}

@ARTICLE{Ianniruberto2020-ap,
  title     = "Origin of shear thinning in unentangled polystyrene melts",
  author    = "Ianniruberto, Giovanni and Marrucci, Giuseppe",
  journal   = "Macromolecules",
  publisher = "American Chemical Society (ACS)",
  volume    =  53,
  number    =  4,
  pages     = "1338--1345",
  year      =  2020,
  doi       = "10.1021/acs.macromol.9b02330",
}

@ARTICLE{Psevdos2026-iq,
  title     = "Flow-induced friction reduction in shear flows of unentangled
               polymer melts",
  author    = "Psevdos, Christos and Ianniruberto, Giovanni",
  journal   = "Macromolecules",
  publisher = "American Chemical Society (ACS)",
  volume    = 59,
  number    = 16,
  pages     = "9154--9161",
  year      =  2026,
  doi       = "10.1021/acs.macromol.6c01365",
}

@BOOK{Doi1988-cc,
  title     = "The theory of polymer dynamics",
  author    = "Doi, M and Edwards, S F",
  publisher = "Clarendon Press",
  year      =  1988,
}

@BOOK{Bird1987-la,
  title     = "Dynamics of polymeric liquids, volume 2: Kinetic Theory",
  author    = "Bird, R B and Curtiss, C F and Armstrong, R C and Hassager, O",
  publisher = "Wiley",
  year      =  1987,
}

@ARTICLE{Rouse1953-og,
  title     = "A theory of the linear viscoelastic properties of dilute
               solutions of coiling polymers",
  author    = "Rouse, Jr, Prince E",
  journal   = "J. Chem. Phys.",
  publisher = "AIP Publishing",
  volume    =  21,
  number    =  7,
  pages     = "1272--1280",
  year      =  1953,
  doi       = "10.1063/1.1699180",
}

@BOOK{Larson1988-nn,
  title     = "Constitutive equations for polymer melts and solutions",
  author    = "Larson, Ronald G",
  publisher = "Butterworth-Heinemann",
  year      =  1988,
  doi       = "10.1016/c2013-0-04284-3",
}

@ARTICLE{Cox1958-vy,
  title     = "Correlation of dynamic and steady flow viscosities",
  author    = "Cox, W P and Merz, E H",
  journal   = "J. Polym. Sci.",
  publisher = "Wiley",
  volume    =  28,
  number    =  118,
  pages     = "619--622",
  year      =  1958,
  doi       = "10.1002/pol.1958.1202811812"
}

\end{document}